\documentstyle[11pt]{article}
\date{}
\begin{document}
\sloppy
\author{V. Majern\'{\i}k,\\
Department of Theoretical Physics, Faculty of Science,
Palack\'y University,\\
T\v r. 17. listopadu 50, CZ-77207 Olomouc, Czech Republic}
\title{Cosmological models with $\Omega_M$-dependent cosmological
constant}

\maketitle

tel: 00420/68/5634278, fax: 00420/68/5225246, e-mail:
majerv@prfnw.upol.cz

\begin{abstract}
We investigate the evolution of the scale factor in a
cosmological model in which the cosmological constant is given by the scalar
arisen by the contraction of the
stress-energy tensor.
\end{abstract}

\section{Introduction}

A revolutionary development seems to take place in cosmology during the last
few years. The evidence continues to mount that the expansion of the
universe is accelerating  rather than slowing down.
New observation suggests a
universe that is leight-weight, is accelerating, and is flat
\cite{PER} \cite{pe} \cite{BO}.
To induce cosmic acceleration
it is necessary to consider some
components, whose equations of state are different from baryons,
neutrinos, dark matter, or radiation considered in the standard
cosmology.

As it is well-known, one way to account for cosmic acceleration is the introduction a new type
of energy, the so-called {\it quintessence} ("dark energy"), a dynamical,
spatially inhomogeneous form of energy with negative pressure \cite{SH}. A common
example is the energy of a slowly evolving scalar field with positive
potential energy, similar to the inflation field in the inflation
cosmology.
The quintessence cosmological scenario (QCDM) is a
spatially flat FRW space-time dominated by the radiation at early times,
and cold dark matter (CDM) and quintessence (Q) later time.
A series of papers of Steinhardt et al. is devoted to
the various quintessence cosmological models \cite{ST} (a number of follow-up
studies are underway). The
quintessence is supposed to obey an equation of state of the form
\begin{equation} \label{1}
p_Qc^{-2}=w_Q\varrho_Q,\qquad
-1<w_Q<0.
\end{equation}
In many models $w_Q$ can vary over time. For the vacuum energy (static cosmological
constant), it holds $w_Q=-1$ and $\dot w_Q=0$.

In what follows we present
a variant of the quintessence cosmological scenario in which the content of
black energy is given by the cosmological constant.
Like that of many other features of relativistic cosmology, history of
the static and dynamic cosmological constant in Einstein's equations abounds in
peculiarities and paradoxes.
The question is of fundamental
significance in present-day cosmology and its discussion raises
fundamental issues in
the interpretation of cosmical constant itself.
The possible existence of very small but non-zero cosmological constant
revives in these days due to new observation in cosmology.

Due to this fact, there are many phenomenological ansatzes for the
cosmological constant more or less justified by physical arguments (see,
e.g. [19]). We remark that observational data indicate that $\lambda\approx 10^{-55} cm^{-2}$ while
particle physics prediction
for $\lambda$ is greater than this value by factor of order $10^{120}$.
This discrepancy is known as the {\it cosmological constant problem}.
The vacuum energy assigned to $\lambda$
appears very tiny but not zero. However, there is no
really compelling dynamical explanation for the smallness of the vacuum
energy at the moment \cite{Is}
(simple quantum-mechanical calculations yield
the vacuum energy much larger \cite{WW}).

A positive non-zero  cosmological constant helps overcome the age problem, connected on the one side
with the hight estimates of the Hubble parameter and with the age of globular
clusters on the other side.
To explain this apparent discrepancy the point of view has often been adopted
which allows the cosmological constant to vary in time. The idea is that during the
evolution of universe the "black" energy linked with cosmological constant
decays into the particles causing its decrease.

As it is well-known, the Einstein field equations with a non-zero $\lambda$ can be rearranged so
that their right-hand sides consist of two terms:
the stress-energy tensor of the ordinary
matter and an additional tensor
\begin{equation}  \label{2}
T^{(\nu)}_{ij}=\left (\frac{c_4\lambda}{8\pi G}\right)
g_{ij}=\Lambda g_{ij}.
\end{equation}
In common discussions, $\Lambda$ is identified with vacuum energy
because this quantity satisfies the requirements asked from $\Lambda$, i.e.
(i) it should have the dimension of energy density, and (ii) it
should be invariant under Lorentz transformation. The second property
is not satisfied for arbitrary systems, e.g. material systems and
radiation. Gliner \cite{G} has shown that the
energy density of vacuum represents a scalar function of the
four-dimensional space-time coordinates so that it satisfies both above
requirements. This is why  $\Lambda$ is commonly {\it identified}
with the vacuum energy.

However, there may be generally other quantities
satisfying also the above requirements.
Instead of identifying $\Lambda$ with the vacuum energy
we have identified $\Lambda$ in \cite{MAL}
with the stress-energy scalar $T=T^i_i$ a scalar
which
arises by the contraction of the
stress-energy tensor of the ordinary matter $T^{j}_i$.
This quantity {\it likewise} satisfies both above requirements, i.e.,
it is Lorentz invariant and has the dimension of the energy density.
Hence, we make the ansatz
\begin{equation} \label{3}
\Lambda_A=\frac{c^4 \lambda_A} {8\pi G}= \kappa T^i_i= \kappa T
\end{equation}
or
\begin{equation} \label{4}
\lambda_A=\frac{8\pi G\kappa T}{c^4},
\end{equation}
where $\kappa$ is a dimensionless constant to be determined.
$\Lambda_A$ is a dynamical quantity, often changing over time, representing, in the
quintessence theory, the quintessence component.
In contrast with some other cosmological models, we
suppose that the universe consists of a mixture of the {\it ordinary} mass-energy
and the quintessence component functionally linked with $T$ via the
cosmological constant $\lambda_A$.
We note that there are similar attempts to identify $\lambda$ with
the Ricci scalar (see \cite{AL}).

In what follows, we introduce a cosmological model with
the additional cosmical term $\Lambda=\kappa T$. The constant $\kappa$
is specified by the assumption that the energy density of the universe
is equal to its critical value.
We show that in the matter-dominated universe the evolution of the scale factor in this model is in
matter-dominated universe
determined by the density parameter $\Omega_M$ in a relatively simple way.

\section{Friedmann's model with a $\Omega_M$-dependent cosmological
constant}

The standard Einstein field equations are (see, e.g. \cite{W})
\begin{equation} \label{5}
R_{ij}-g_{ij}(1/2)R-\lambda g_{ij}=
\frac{8\pi G} {c^4} T^{(m)}_{ij}.
\end{equation}
These equations can be rewritten in the form
\begin{equation}
\label{6}
R_{ij}-g_{ij}(1/2)R=\frac{8\pi G}{c^4}(T^{(m)}_{ij}+T^{(v)}_{ij}),
\end{equation}
where
$$T^{(v)}_{ij}=g_{ij}\Lambda\quad\Lambda={\lambda c^4\over
8\pi G}.$$
Putting $\Lambda=\Lambda_A=\kappa T$ we have
$$T^{(v)}_{ij}=g_{ij}\kappa T.$$
and Eq.(\ref{6}) becomes
\begin{equation}
R_{ij}-g_{ij}(1/2)R = \frac{8\pi
G}{c^4}\left [T^{(m)}_{ij}+g_{ij}\kappa T\right].
\label{7}
\end{equation}
In a homogeneous and isotropic universe characterized by the
Friedmann-Robertson-Walker line element the Einstein equations with
matter in the form of a perfect fluid
and non-zero cosmical term $\lambda$ acquire the following form
\begin{equation} \label{8}
{3\dot R(t)^2\over R(t)^2}= 8\pi G\rho +\lambda
c^2-3\frac{kc^2}{R^2(t)},
\end{equation}
and
\begin{equation}                       \label{9}
\ddot R(t)= \frac{4\pi G}{3}(-\rho-3p/c^2)+\frac{\lambda c^2}{3}R(t),
\end{equation}
where $R(t)$ is the time-dependent scale factor.

To determine the exact form of $\Lambda_A$ which is to be inserted in
Eqs.(8) and (9) we have to specify $\kappa$ and $T$.
$T$ can be derived from the tensor $T^{i}_{j}$ and
$\kappa$ in Eq.(\ref{3}) we determine by assuming that
the universe is flat, i.e., $\Omega_{tot}=1$. This is consistent with
the inflationary cosmology which assumes that the universe is spatially flat and
that its total energy density is equal to the critical
density $(\Omega_{tot}=1)$. This assumption is also conformed
by the current measurement of the
cosmic microwave background anisotropy [18]. Since $\Omega_M<1$ we
suppose that the remaining energy required to produce a geometrical flat
universe is given by the equation
$$\Omega_M+\Omega_Q=\Omega_M+\kappa \Omega_M=1.$$
This gives
\begin{equation} \label{11}
\kappa=\frac{1}{\Omega_M}-1.
\end{equation}
By specifying $\kappa$ and $T$, the cosmological constant $\Lambda_A$
is uniquely determined so we can
investigate the cosmological models with $\Lambda_A$ for the different
values of
$\Omega_M$.
\begin{equation} \label{S}
\Lambda_A=\left (\frac{1}{\Omega_M} -1\right ) T.
\end{equation}

The stress-energy tensor of the cosmic medium $T^{i}_{j}$ in the
everywhere local rest frame has only four non-zero components
$T^0_{0}=\varrho c^2, T^1_1=T^2_{2}=T^3_3=-p$ \cite{UL}. Therefore,
\begin{equation} \label{T}
T=\varrho c^2-3p/c^2.
\end{equation}
Setting $p=w\rho c^2$ we have
\begin{equation}  \label{TT}
T=\rho c^2(1-3w).
\end{equation}
We see that $\Lambda_A$ is a function of both the mass-energy density
and some {\it stress} components.

A comprehensive analysis of Eqs.(\ref{8}) and (\ref{9}) has been carried out in
\cite{FI} for static cosmological term, and in \cite{OC} for a varying
cosmological term.
A quantitative analysis of solutions to Eqs.(\ref{8}) and
(\ref{9}) can be
gained by eliminating $\rho $ in these equations and combining them into
a single equation for the evolution of the scale factor in the presence
of a $\lambda$-term \cite{SA}
\begin{equation}    \label{11'}
\frac{2\ddot R}{R}+(1+3w)(\frac{\dot R^2}{R^2}+\frac{kc^{2}}{R^2})-(1+w)\lambda
c^2=0,
\end{equation}
\section{Matter-dominated epoch}
In what follows, we consider the matter-dominated and flat
universe, i.e. we set
$w=0$ and $k=0$. In this universe $T=\rho_M c^2$.
Inserting $T=\rho_M c^2$ in the equation for $\Lambda_A$ we have
\begin{equation}  \label{13}
\Lambda_A=\kappa \varrho_Mc^2 =\left (\frac{1}{\Omega_M}-1\right
)\rho_M c^2
=(\varrho_{crit}-\varrho_M.)c^2=\varrho_{crit}(1-\Omega_M)c^2,
\end{equation}
The critical density $\rho_{crit}$ we obtain
by inserting Eq.(\ref{13}) into Eq.(\ref{8})
\begin{equation} \label{O}
\rho_{crit}=\frac{3\dot R^2}{R^2}.
\end{equation}
The insertion of Eq.(\ref{O}) into Eq.(14) yields immediately
the equation for the evolution of $R(t)$ in the matter dominated epoch
\begin{equation} \label{14}
\ddot R(t)= \left (1-\frac{3}{2}\Omega_M(t) \right) \frac{(\dot
R(t))^2}{R(t)}.
\end{equation}
In our model, this equation describes
the time dependence of the scale factor as a
function of $\Omega_M(t)$ and represents so the basic equation for the
evolution dynamics of a pressure-free and flat universe.

The exact solution of (\ref{14}) can be found for an arbitrary time function
$\Omega_M(t)$. With the ansatz $R=\exp(y)$ we have
$$ \dot R=\dot y\exp(y), \qquad \ddot R= (\ddot y+(\dot
y)^2)\exp(y)$$
which inserting into Eq.(\ref{14}) yields
$$-(2/3)\Omega_M(t)(\dot y)^2=\ddot y.$$
By putting $\dot y =q$, this equation becomes
the form
$$-(2/3)\Omega_M(t)=\frac{\dot q}{q^2},$$
the solution to which is
$$q=\frac{1}{\int (2/3)\Omega_M(t) dt +C_1}.$$
Since $\dot y =q$ we have
$$y=\int \left(\frac{1}{\int (2/3)\Omega_M(t) dt+C_1}\right) dt +C_2.$$
With $y(t)$, the general solution of Eq.(\ref{14}) is
\begin{equation} \label{15}
R(t)=\exp \int\left (\frac{1}{\int (2/3)\Omega_M(t) dt+C_1}\right) dt +C_2.
\end{equation}
We see that evolution of $R(t)$
in a pressure-free medium is
a function of the time dependence of $\Omega_M(t)$ and the integration
constants $C_1$ and $C_2$.

\section{Evolution of R(t) for the different density parameters}

We present solutions of Eq.(\ref{14}) for some selected constant
values of $\Omega_M$ and
derive then the cosmological parameters of the corresponding models of
universe.\\
(A) For $\Omega_M=0$, i.e. for a massless universe, we get a typical
inflationary solution of Eq.(\ref{14})
$$R(t)=\exp (C_1(t-C_2)).$$
(B) For $\Omega_M=2/3$, the evolution of R(t) is
$$R(t) =C_1+tC_2.$$
With $C_1=0$, it gives
$$R(t)=C_2t.$$
(C) For $\Omega_M=1/3$, the evolution of $R(t)$
is
$$ R(t)=\frac{1}{4}C_1(t^2-2tC_2+C_2^{2})^{1/3}.$$
With $C_2=0$, it gives
$$R(t)=\frac{1}{4}C_1t^2.$$
(D) For $\Omega_M=1$, the evolution of $R(t)$ is
$$R(t)=\frac{C_1^{2/3}(9t^2-18t(C_2+9C_2^2))^{1/3}}{2^{2/3}}.$$
With $C_2=0$, we obtain
$R(t)=Kt^{2/3}$, i.e. the evolution law which is identical with that of the Standard
Cosmology in a pressure-free cosmic medium. We see that in all cases (except A)
$R(t)$ satisfies the initial condition $R(0)=0$
and represents a smoothy increasing functions of
time.\\
We now analyse the cases $A,B,C,D$ in more details.\\
{\bf Case A}. It is tempting to choose for the early universe
 $\Omega_M=0$,
 i.e. to suppose that the universe started in a massless
state and its mass content was created later through the decay of the cosmical term.
Under this assumption we have
\begin{equation} \label{X}
R(t)=\exp(C_1(t-C_2))=R_0\exp (C_1t),\qquad C_1=\frac{1}{t_0}.
\end{equation}
The natural measures for length and time in cosmology is the Planck
length and time, i.e., $l_p=(Gh/c^3)^{1/2}=4.3.10^{-35} \rm
{m}$ and $t_p=(Gh/c^5)^{1/2}=1.34.10^{-43}\rm {s},$ respectively.
It is reasonable to assume that at the very beginning of the cosmic
evolution the radius of the universe was of the order of the Planck
length,
therefore we put in Eq.(19) the integration constant $R_0$ and
$C_1$ equal to $l_p$ and
$1/t_p$, respectively.
Then, we get for
the initial radius and the velocity the values
$$R(0)=l_p=4.3.10^{-35}m.\qquad\dot R(0)=\frac{l_p}{t_p}=c=3.10^{8}
ms^{-1},$$
respectively. The most interesting thing of the A-type universe is its
inflationary character
(for a recent review see \cite{LI}).

{\bf Case B}. In the B-type universe the relevant cosmic parameters are\\
$$ R=C_2t,\qquad \dot R(t)=C_2,\qquad H=\frac{1}{t}\qquad q=0.$$
The age of this universe, if taking $H=50 km s^{-1} Mpc^{-1}$, is
$$t_0=\frac{1}{H_0}=2.10^{10} yr.$$
This age is larger than that in the Standard Model.
However, this universe is not accelerating so it seems not to be compatible with the recent
data.\\
{\bf Case C}.
The relevant cosmological parameters here are
$$R(t)=(1/4)C_1t^2\qquad \dot R=\frac{C_1t}{2}, \qquad q=-\frac{1}{2},\qquad
H(t)=\frac{2}{t}\qquad
\lambda_A=\frac{8}{c^2t^2}.$$
In the C-type universe there is no age problem because $t_0=2/H_0$.
The age of this universe is approximately $4.10^{10} yr$, i.e. old enough for the
evolution of
the globular clusters. This universe is accelerated ($q=-0.5$) .
Its density parameter is $\Omega_M =1/3$
which
corresponds to the recent data.
In this universe the proper distance $L(t)$ to the horizon, which is the linear
extent of the causally connected domain, diverges
$$L(t)=R(t)\int_0^t\frac{d\tau}{R(\tau)}=C_2t^2[-(\frac{2}{\tau})|^{t}_0]
=-\infty, $$
In \cite{PS} is shown that the only way to make the whole of the
observable universe causally connected is to have a model with infinite $L(t)$
for all $t>0$, i.e.
the whole C-type observable universe is causally connected.
It is noteworthy that the decay law for the cosmical constant of the
form
$\lambda=at^{-2}$, was
phenomenologically set by
several authors
whereby different authors used different
physical arguments for its justification
\cite{B}-
\cite{ARB}.
(For a recent review see \cite{OC}).

{\bf Case D}. In this universe we have
$$R(t)=Kt^{2/3} = \frac{C_1^{2/3}9}{2^{2/3}}t^{2/3}.$$
All other parameters of the D-type
universe are identical with
those of Standard Model.

In order to vanish
the covariant divergence of the right-hand side of Eq.(6)
the matter is created
along with energy and momentum.
Therefore, the cosmological constant $\lambda_A$ decays during the cosmological
time and new particles are created. The present rate of mater creation
in the matter dominated epoch is very small [29]
$$n=\frac{1}{R^3}\frac{d(\rho R^3)}{dt}|_0.$$

\section{Final remark}

Summing up we can state:\\
(i) In previous sections we have shown that the density parameter
$\Omega_M$  determines, in our model of the universe,
its entire evolution dynamics. In the basic dynamical
equation (17) the energy density does not explicitly appear only in the
density parameter $\Omega_M$.
We note that the density parameter $\Omega_M$ as the ratio of $\rho_M$
and $\rho_{crit}$ may be {\it finite} although both quantities are infinite.

(ii) There is growing observational evidence
that the total matter of the universe is significantly less than the
critical density. Several authors \cite{OS} \cite {KT}
\cite{TW} have found that the best and simplest fit is provide by
($h=0.65\pm 0.15$)
$$\Omega_M = \Omega_{CDM}+ \Omega_{baryon}\approx [0.30\pm 0.10]
+[0,04+\pm 0.01]\approx 1/3$$
which is approximately the density parameter considered in the C-type universe.
\cite{REI}.
(iii) In the recently popular $\Lambda$CDM cosmological model, which
consists of a mixture of vacuum energy and cold dark matter, a serious
problem exists called in \cite{ST} as the cosmic  coincidence problem.
Since the vacuum energy density is constant over time and the matter
density decreases as the universe expands it appears that their ratio
must be set to immense small value ($\approx 10^{-120}$) in the early universe
in order for the two densities to nearly coincide today, some billions
years later. No coincidence problem exists in the C-type universe
because $\Lambda_A$ here is functionally connected with
$\Omega_M$ in such a
way that this ratio in the matter dominated epoch does not vary over
time.

(iv) In the radiation dominated epoch $w=1/3$ and, according to
Eq.(\ref{TT}, $T=0$. The evolution dynamics
in this epoch runs so as if $\lambda =0$.
 
In conclusion, when comparing the cosmological parameters of the different cosmological
models we see
that the recent observational data of the flat and acceleration universe
are most consistent with the C-type universe.
This universe is
leight-weight, is strictly flat, is accelerating, is old enough and is causally
connected.

\end{document}